# Optimizing Precision for Open-World Website Fingerprinting

Tao Wang
*Department of Computer Science and Engineering*
*Hong Kong University of Science and Technology*
taow@cse.ust.hk

**Abstract**

Traffic analysis attacks to identify which web page a client is browsing, using only her packet metadata — known as website fingerprinting — has been proven effective in closed-world experiments against privacy technologies like Tor. However, due to the base rate fallacy, these attacks have failed in large open-world settings against clients that visit sensitive pages with a low base rate. We find that this is because they have poor precision as they were designed to maximize recall.

In this work, we argue that precision is more important than recall for open-world website fingerprinting. For this reason, we develop three classes of *precision optimizers*, based on confidence, distance, and ensemble learning, that can be applied to any classifier to increase precision. We test them on known website fingerprinting attacks and show significant improvements in precision. Against a difficult scenario, where the attacker wants to monitor and distinguish 100 sensitive pages each with a low mean base rate of 0.00001, our best optimized classifier can achieve a precision of 0.78; the highest precision of any known attack before optimization was 0.014. We use precise classifiers to tackle realistic objectives in website fingerprinting, including selection, identification, and defeating website fingerprinting defenses.

## 1 Introduction

The massive scale of state-level surveillance was affirmed in 2013 by the Snowden revelations. Yet, it has continued to expand. Recent events show that state actors continue to take an interest in large-scale persistent and passive monitoring of the Internet. In November 2016, the UK passed the Investigatory Powers Act, described by the executive director of the Open Rights Group as "unprecedented powers to monitor and analyze UK citizens' communications regardless of whether [they] are suspected of any criminal activity." [17] A similar growth in surveillance can be seen in many countries.

Individuals seeking privacy in web browsing may need to use privacy enhancing technologies such as VPNs and anonymity networks like Tor. These technologies encrypt and redirect the user's traffic across proxies, hiding sender, recipient, and packet contents from a passive observer. However, they fail to hide significant features of user traffic, such as packet frequency, timing, order, and direction. **Website Fingerprinting** (WF) attacks use these features to identify the client's destination web page, compromising her privacy. WF attacks only use metadata, which can be monitored and recorded freely in many countries as they are not considered to be personal, legally [14].

The WF attacker wishes to monitor a set of sensitive web pages and identify when a client visits these pages. In the closed-world scenario, the classifier is tested only on the same sensitive web pages that it is trained on. In the harder open-world scenario, the classifier must also correctly identify which pages it has never seen before and are therefore not sensitive.

WF attacks have been proven effective against real-world privacy technologies in the closed-world scenario [3, 10, 15, 21, 28]. However, there is a general academic consensus that WF fails in the open world due to the base rate fallacy. In 2016, Panchenko et al. studied WF attacks in a large open world and concluded that "no existing [WF] method scales when applied in realistic settings [in the open world]" [21]. This paper tackles the unsolved problem of **open-world website fingerprinting** (OWF). The open-world scenario has been challenging for other fields that rely on machine learning as well, including forensic analysis [24], intrusion detection systems [23], and medical imaging [26]. Open-world failure in forensic analysis has been subject to public controversy [19] as it leads to wrongful convictions.

We argue that the reason why WF fails in the open world is because previous work has focused on the wrong metric. All known WF attacks attempt to maximize *recall* (true positive rate), as is standard in closed-world evaluation. However, a successful open-world classifier must have high *precision* to avoid the base rate fallacy. As this argument is central to our work, we expound our point in Section 2. A known method to adapt closed-world classifiers for OWF is to add a new class that represents the open world, but we will show that this method is often both inaccurate and imprecise.

Table 1: How we count the number of true positives ($N_{TP}$), wrong positives ($N_{WP}$), and false positives ($N_{FP}$). After counting them, we obtain the true positive rate $R_{TP} = N_{TP}/N_P$, wrong positive rate $R_{WP} = N_{WP}/N_P$, and false positive rate $R_{FP} = N_{FP}/N_N$.

| | | Classified as | | |
|---|---|---|---|---|
| | | Correct sensitive class | Wrong sensitive class | Non-sensitive class |
| True class | Sensitive class (# = $N_P$) | True Positive (# = $N_{TP}$) | Wrong Positive (# = $N_{WP}$) | False Negative (# = $N_P - N_{TP} - N_{WP}$) |
| | Non-sensitive class (# = $N_N$) | Not possible | False Positive (# = $N_{FP}$) | True Negative (# = $N_N - N_{FP}$) |

This work contributes three classes of techniques that can improve the open-world precision of any closed-world classifier; we call them Precision Optimizers (POs). We demonstrate the effectiveness of our POs by significantly improving the precision of six known WF attacks. We will find that even relatively simple POs can significantly improve the precision of WF attacks, demonstrating incorrect optimization in WF attacks. We present these results in Section 3; our best PO can improve a classifier's precision from 0.014 to 0.78 in a scenario with a very low base rate.

We show that our improved classifiers are better able to handle realistic scenarios — and thus pose a greater threat to privacy — in Section 4. In particular, we use them to attack several WF defenses in Section 4.3, and show that we can significantly improve precision against defenses as well. We discuss several pertinent WF issues in Section 5, give a survey of related work in Section 6, and conclude in Section 7.

## 2 Background

### 2.1 Terminology and Setting

A classifier takes as input a testing element and determines which class it belongs to. In our case, the testing element is a sequence of packets (with timing, size and direction) and each class is a web page.[1] When the classifier claims that the tested packet sequence is sensitive, it is known as a *positive*; it is a *true positive* (TP) if the tested packet sequence came from the same page the classifier identified, and it is a *false positive* (FP) if the tested packet sequence came from a non-sensitive page. We define a *wrong positive* (WP) to be a sensitive page mistaken as another sensitive page. While some previous works have considered a wrong positive to be a false positive, we do not, because the tested packet sequence did not come from a non-sensitive page. Refer to Table 1 for an illustration of these terms and how their rates (TPR/WPR/FPR) are defined. We also refer to TPR as recall; previous WF works used recall as the main metric to compare and optimize WF attacks.

Our WF attacker is a *passive* eavesdropper that is *local* to the client. The attacker watches packet sequences sent by the client, and knows the client's identity. The privacy-sensitive client uses encryption with proxies to hide her packet contents and destination web page from the attacker. The attacker compromises her privacy by using a classifier that decides if the client is visiting a set of *sensitive pages*. The classifier is trained on supervised packet sequence data, which he collected by visiting the sensitive pages himself. The set of sensitive pages is, inevitably, a small subset of all web pages, so the attacker must be able to recognize when the client is not visiting sensitive web pages, avoiding the *base rate fallacy*.

The base rate fallacy describes the following problem: if the FPR is higher than the base rate of positive events (sensitive web page accesses), then the positive classifications of the classifier are mostly false, even if recall is 100%. In this case, a classifier with a high recall may still be useless in practice, as most of its positive classifications are false. To illustrate this with an example, consider a boy who always cries wolf when he spots a wolf (100% recall), but has a 20% chance of crying wolf every day when he does not spot a wolf (20% FPR), out of boredom. If a wolf appears once every 46 days, only 10% of the boy's cries are correct (10% precision), and the villagers may ignore him despite his perfect recall.

The base rate fallacy is the chief challenge of OWF. To avoid the base rate fallacy, the attacker wants to classify a page access as sensitive only if he is *confident* that the classification is correct. In this work, we propose and evaluate three classes of techniques to improve OWF precision, which we call *precision optimizers (POs)*.

---

[1] In this work, we use sequences of Tor cells where each cell has the same size; we still call them packet sequences for generality.

In this work, we also consider WF *defenses*. These defenses have been proposed to defend web-browsing clients against WF attacks. WF defenses are applied by the client and her proxies, and they transform the packet sequence to disrupt the attacker's ability to classify them correctly. We will evaluate the effectiveness of these defenses in the open world with our optimized classifiers.

We present our notations in Table 2.

## 2.2 r-precision

The base rate fallacy shows that the TPR of WF classifiers does not tell the whole story. We must also consider the WPR and FPR as well. A classifier should only be considered effective if its positive classifications are largely true; otherwise, the attacker cannot act on its positive classifications. None of the above metrics, alone, tell us if the classifier satisfies this criteria. We also need to include a fourth metric: the base rate at which the client accesses sensitive web pages. Without considering the base rate, it is not clear how any attack fares against the base rate fallacy.

We propose the use of *r*-**precision** to measure WF attacks, which considers all of the four metrics above. We derive it in the following. The number of positive classifications returned by the classifier is the sum of the numbers of true, wrong, and false positives ($N_{TP} + N_{WP} + N_{FP}$). Among its positive classifications, the ratio of true positive classifications is equal to:

$$\pi = \frac{N_{TP}}{N_{TP} + N_{WP} + N_{FP}}$$

This is the *precision* of the classifier. In an experimental setting, we measure rates ($TPR$, $WPR$, $FPR$) rather than direct counts, which requires an explicit definition of the base rate.

$$\pi_r = \frac{R_{TP} \cdot N_P}{R_{TP} \cdot N_P + R_{WP} \cdot N_P + R_{FP} \cdot N_N}$$
$$= \frac{R_{TP}}{R_{TP} + R_{WP} + \frac{N_N}{N_P} \cdot R_{FP}}$$

Setting $r = N_N/N_P$, we arrive at our formulation of *r-precision* that incorporates the base rate.[2] *r*-precision is the analytical basis used in the remainder of this paper. *r* (which we call the base ratio) is the relative likelihood of negative events (client visiting any non-sensitive page) to positive events (client visiting any sensitive page). A higher *r* reduces *r*-precision holding all other rates constant, making the classification problem harder.

Our objective is to maximize $\pi_r$ for WF attacks. In our experiments, we present results for $r = 10$ and $r = 1000$, representing respectively an easy and hard classification setting. Note that *r* describes the total base ratio of all sensitive pages (we consider 100 sensitive pages in our work). For example, if the attacker wants to monitor 100 sensitive pages and the client has a 1/100,000 chance of visiting each sensitive page, the attacker's precision would be correctly captured by our $\pi_{1000}$ scenario.

[2]The base rate is in fact $\frac{N_P}{N_N + N_P} = \frac{1}{1+r}$.

Table 2: Notation used in this paper.

| | |
|---|---|
| $N_P, N_N$ | # of positives, negatives |
| $N_{TP}, N_{WP}, N_{FP}$ | # of true, wrong, false positives |
| $R_{TP}, R_{WP}, R_{FP}$ | True, wrong, false positive rates |
| $r$ | Base ratio, equal to $N_N/N_P$ |
| $\pi_r$ | precision for base ratio $r$ |
| PO | Method to improve WF precision |
| Recall | Equal to $R_{TP}$ |
| P | A packet sequence |
| C | A class |

*r*-precision is more intuitive for the open world than TPR, WPR and FPR. *r*-precision is equal to the probability that the WF classifier is correct when it says that the client has visited a sensitive page, if the client visits *r* times as many non-sensitive pages as sensitive pages. Another advantage to using *r*-precision is that it directly considers the base rate. In several previous works [21, 22], researchers noted that the performance of their WF classifiers dropped when they considered a larger open world, which led them to conclude that OWF was not realistic. This is an incomplete conclusion because they also indirectly increased *r* when expanding the open world. We will observe that *r*-precision does not decrease with larger open world sizes (in fact, it increases to a point because the classifier is trained on more information), so we have avoided this issue.

The minimum recall we accept in this work is 0.2. Our approach to OWF is novel compared to previous work; we maximize *r*-precision while ensuring the recall is acceptable, while previous works maximize recall [3, 21, 22, 27–29]. Our new approach avoids the base rate fallacy and better fits the open-world attack model.

## 2.3 Does precision trump recall?

In the classical formulation of WF, the effectiveness of an attack is measured with its recall (TPR). We present three arguments to convince the reader that it is more important to optimize the precision of WF attacks than their recall.

**Base rate fallacy.** Section 2.1 shows that a WF classifier can never be said to have avoided the base rate fallacy no matter how high its recall is. There are billions of pages on the Internet, and the base rate for the vast majority of pages is low. Furthermore, pages change over time, and dynamic content further distorts page information [12, 30]. Only if the attacker is confident that the classifier's positive classifications are true (precision is high) can the attacker act on them. The base rate fallacy highlights the epistemological deficiency of recall in the open world: it does not tell us anything about open-world effectiveness.

**Chilling effect.** Web-browsing clients using anonymity networks are sensitive to privacy and do not want the attacker to capture any of their browsing behavior. When WF precision is high, even a moderate recall may trouble privacy-sensitive users. For example, a whistleblower who would suffer a 10% recall — a 10% chance of revealing each sensitive page access to any eavesdroppers — may instead choose to self-censor or use a WF defense (which are not yet implemented and may incur significant overhead).

**Repeated visits.** Browsing patterns are often consistent and self-repeating, and users visit the same websites frequently. Repeatedly visiting the same page would allow a low-recall, high-precision attacker to observe it. Even with a low recall, the WF attacker would eventually be able to determine if the client is interested in particular sensitive pages. A high recall is unnecessary in such a scenario. Repeated visits cause a gradual decay in online privacy.

One way to improve open-world performance is to minimize FPR, but this can be achieved trivially. For any algorithm with TPR $R_{TP}$, WPR $R_{WP}$ and FPR $R_{FP}$, it is possible to achieve TPR $\alpha R_{TP}$, WPR $\alpha R_{WP}$ and FPR $\alpha R_{FP}$ for any $0 \leq \alpha \leq 1$ by simply randomly classifying positive instances as negative instances with probability $\alpha$. This procedure, which produces the trivial diagonal ROC curve, would have no effect on precision. Since we want to increase the precision of classifiers in our work, we must develop better techniques.

### 2.4 Experimental setup

We collected our data set for OWF between March and June 2017 with Tor Browser 6.5.1 on Tor 0.2.9.10. We focus exclusively on Tor because it is both popular and relatively resilient. Furthermore, it is the most difficult for WF out of currently usable anonymity networks [10], WF attacks designed to succeed in the Tor scenario also tend to succeed against other privacy technologies [27].

We collected a set of sensitive (monitored) pages and non-sensitive (non-monitored) pages. We chose the top 133 pages on Alexa as the sensitive set, visiting each of them 600 times, and the next 60,000 pages on Alexa as the non-sensitive set, visiting each of them once. We decided that the sensitive pages should be top sites to best represent a practical client's possible interests. We visited each site for one minute, collecting all cells. After filtering away pages which frequently did not load, we ended up with 100 pages (500 instances each) in the sensitive set and 50,000 pages in the open-world set.[3]

For each web page, we collected the times, sizes, and directions of all packets, from which Tor cells can be derived [29], representing a local, passive attacker's information.

For convenience, we may describe a subset of our full data set using numeric notation such as 50x100+20, which denotes that the subset has 50 monitored pages, 100 instances of each page, and 20 non-monitored pages (non-monitored pages always have one instance each).

In this work, we measure the 95% confidence interval of a statistic $\hat{x}$ by taking:

$$C(\hat{x}) = 1.96 \sqrt{\frac{\hat{x}(1-\hat{x})}{n}}$$

This is the confidence interval of the mean for the normal distribution using the Wald method, and we apply it to TPR, WPR, and FPR. We then write $\hat{x} \pm C(\hat{x})$ to show the confidence interval of $\hat{x}$. We are able to use the normal distribution as our $n$ is large (usually 50,000). However, the above does not apply to $r$-precision. Recall that the definition of $r$-precision is:

$$\pi_r = \frac{R_{TP}}{R_{TP} + R_{WP} + r \cdot R_{FP}}$$

Let us denote $R_{TP}^{max} = R_{TP} + C(R_{TP})$, and $R_{TP}^{min} = R_{TP} - C(R_{TP})$, and correspondingly for $R_{WP}$ and $R_{FP}$. We take a naïve 95% confidence interval by computing the maximum $\pi_r$:

$$\pi_r^{max} = \frac{R_{TP}^{max}}{R_{TP}^{min} + R_{WP}^{min} + r \cdot R_{FP}^{min}}$$

We take $C(\pi_r) = \pi_r^{max} - \pi_r$ and show the confidence interval as $\pi_r \pm C(\pi_r)$.[4] When $r$ is large (as in our experiments), precision is often dominated by the $r \cdot R_{FP}^{min}$ term in the denominator.

When $\pi_r$ is *high*, $C(\pi_r)$, its confidence interval, is unstable. This is because a high $r$-precision indicates a very small number of false positives, especially in the $r = 1000$ scenario. For example, consider an experiment with $N_P = 50000$ and $N_N = 50000$ where we find that $R_{TP} = 0.36, R_{WP} = 0, R_{FP} = 0.00004$. This gives $\pi_{1000} = 0.9$. However, this means that $N_{FP} = R_{FP} \cdot N_N = 2 < 10$, which fails the basic criteria for using the normal distribution as the confidence interval. In such cases, we calculate the maximum FPR using the Wilson method (which better suits the extremely small rate $R_{FP}$), with $z = 1.96$ for the 95% confidence interval:

$$R_{FP}^{max} = \frac{R_{FP} + \frac{z^2}{2N_N} + z\sqrt{\frac{R_{FP}(1-R_{FP})}{N_N} + \frac{z^2}{4N_N^2}}}{1 + \frac{z^2}{N_N}}$$

---

[3]Most of the 33 pages that failed to load in the top pages were Chinese, which usually cannot be accessed from Tor exit nodes.

[4]This $C(\pi_r)$ is larger than an alternative $C'(\pi_r) = \pi_r - \pi_r^{min}$, so the estimation is cautious.

Then, we take the minimum precision as

$$\pi_r^{min} = \frac{R_{TP}^{min}}{R_{TP}^{max} + R_{WP}^{max} + r \cdot R_{FP}^{max}}$$

Finally, we express the precision with its lower bound of $\pi_r \geq \pi_r^{min}$. We present no value for the upper bound of precision as it cannot be accurately measured. Continuing the above example, we find $R_{TP}^{min} = 0.356$, $R_{TP}^{max} = 0.364$, $R_{WP}^{max} = 0$, and $R_{FP}^{max} = 0.00015$, so we would write $\pi_{1000} \geq 0.70$, not $\pi_{1000} = 0.9$. We use the Wilson method for $\pi_{1000}$ whenever $N_{FP} < 10$.

### 2.5 WF attacks

We choose a subset of WF attacks to test our POs. All chosen attacks were state-of-the-art at the time they were published, and each attack achieved success in closed-world website fingerprinting. We summarized the attacks briefly in Table 3, in order of publication; we describe these attacks in more detail in the Appendix.

In previous works, some of the above closed-world classifiers were adapted for OWF simply by adding an open-world class representing non-sensitive pages. As we will see in Section 3.1, this technique achieves poor results in practice. We need to come up with better techniques to optimize precision.

## 3 Precision Optimizers

We modify closed-world classifiers to achieve high $r$-precision for OWF using Precision Optimizers (POs). Our POs teach the underlying classifier to be conservative, such that it would assign a packet sequence to the negative class (non-sensitive web page) if it is not certain about its classification. This reduces FPR, thus increasing $r$-precision.

To optimize $r$-precision, we first ask the closed-world classifier to classify the element as usual. If it is a negative classification, we do not apply PO. Otherwise, if the classifier decides the element should be classified as a sensitive page (we refer to that class as the *assumed class*), we ask a PO whether or not we should *reject classification* of the assumed class. The PO may agree with the assumed class, or it may reject the assumed class and instead classify the element as non-sensitive. (Our POs do not change a sensitive classification to another sensitive classification.) This is shown in Figure 1.

Our POs are designed to be *classifier-agnostic*: they treat the classifier as a black box and they can be applied to all classifiers. Some POs are also parametrically tuneable to allow maximization of $r$-precision.

We start this section by motivating our work with an experiment on the baseline $r$-precision of WF classifiers without any POs (Section 3.1). Then, we present three types of POs: confidence-based POs (Section 3.2), distance-based POs (Section 3.3), and ensemble POs

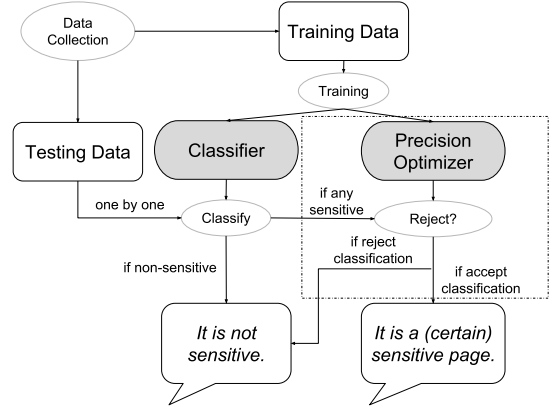

Figure 1: Flowchart describing how we classify an element. Our Precision Optimizers (POs) modify classification to improve precision as represented by the dotted box.

(Section 3.4). Our techniques are inspired by techniques used in clustering and ensemble learning.

### 3.1 Baseline precision

We first present the baseline precision as a comparison before applying any of our POs on the classifiers. We experimented on the classifiers in Table 3, using the experimental setup and methodology described in Section 2.4. Besides the 100 classes each representing a sensitive page, we add a 101st "non-monitored class" with all 50,000 non-sensitive pages. We present the best results for 10-precision and 1000-precision in Table 4.

The above experiment shows that none of the algorithms are precise in both scenarios. This demonstrates that the trivial strategy of adding a non-sensitive class, used in previous work, is unable to achieve high precision. The best-performing attacks are Ca-OSAD and Ha-KFP, though roughly both of their positive classifications are almost half wrong in the easier $\pi_{10}$ scenario (when one-tenth of the user's page visits are sensitive to the attacker).

All 100,000 packet sequences were part of the testing set for each attack except Ca-OSAD.[5] To form the training set, we attempted to use 10-fold cross validation, so that the training set would have 100x450+45000 elements and the testing set would have 100x50+5000 elements, and we would repeat this ten times with ten disparate testing sets. However, we found that this was impossible for three attacks — Ha-KFP, Pa-SVM, and Pa-CUMUL. For the first attack, we ran out of memory. For the others, it was due to poor convergence properties of the underlying SVM classifier. For each of these cases, we reduced each training set from 90,000 elements to 10,000 elements during 10-fold cross validation. (This is approximately the same size or larger than the training sets used in their works.)

---
[5] We tested Ca-OSAD only on 100x50+5000 elements because of the computational time involved to compute the custom SVM distance kernel, which scales with the square of the number of instances. On the full data set, it would've taken around 300,000 CPU hours.

Table 3: Summary of the known WF attacks we will optimize for *r*-precision in this paper.

| Name | Classifier | Classification mechanism |
|---|---|---|
| Bi-XCor [1] | Scoring | Cross correlation on inter-packet timing and lengths |
| Pa-SVM [22] | SVM | Sequence features such as length and packet ordering |
| Ca-OSAD [3] | SVM | Custom kernel using Levenshtein distance |
| Wa-kNN [28] | kNN | Custom-weighted kNN on various sequence features |
| Ha-kFP [9] | Random Forest | Random forest classifier on various sequence features |
| Pa-CUMUL [21] | SVM | Cumulative packet sizes |

Table 4: Baseline 10-precision ($\pi_{10}$) and 1000-precision ($\pi_{1000}$) for six known WF attacks without any PO.

| Name | $\pi_{10}$ | $\pi_{1000}$ |
|---|---|---|
| Bi-XCor [1] | $.059 \pm .001$ | $.00066 \pm .00001$ |
| Pa-SVM [22] | $.249 \pm .006$ | $.0038 \pm .0001$ |
| Ca-OSAD [3] | $.61 \pm .04$ | $.036 \pm .014$ |
| Wa-kNN [28] | $.313 \pm .005$ | $.0048 \pm .0001$ |
| Ha-kFP [9] | $.53 \pm .01$ | $.0137 \pm .0006$ |
| Pa-CUMUL [21] | $.311 \pm .007$ | $.00499 \pm .00016$ |

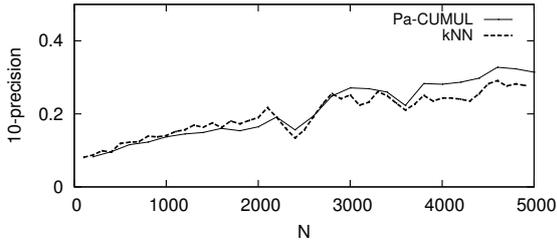

Figure 2: 10-precision for `Wa-kNN` and `Pa-CUMUL` when trained on 100x50+*N* elements, where *N* is the size of the open-world class.

In previous work, it has been shown that a larger open world class size in the training set sometimes decreased classifier performance [21, 22]. This observation has been used to argue that WF attacks do not perform well in the open world [18]. We believe this is the case only because the wrong metric was used. We measured $\pi_{10}$ for `Wa-kNN` and `Pa-CUMUL` on a testing set of 100x50+5000 elements and a training set of 100x50+*N* elements, where we varied *N* (the open world class size) from 100 to 5000. We plotted the results in Figure 2. We see that increasing *N* in fact *increases* precision, as may be expected of a competent classifier. The increase is, however, somewhat uneven; this is partly due to the inherently noisy nature of precision on a relatively small testing set, but it also shows that adding an open world class cannot achieve high precision consistently.

### 3.2 Confidence-based PO

To classify an input element *P*, some classifiers compute some matching function *match* between *P* and all trained classes *C*, and classify *P* to the class that maximizes the value of the function:

$$\underset{C}{\mathrm{argmax}}\, match(P,C)$$

As an example, in the following we describe the *match* function used by Support Vector Machines (SVMs). SVMs are used by several WF attacks [3, 21, 22, 29]. We specify the *match* function for other classifiers in the Appendix.

SVMs are binary classifiers: they seek to find an optimal separator between two classes in training. We denote $f_{C,C'}(P) \in \{C,C'\}$ as the classification output of an SVM trained on two classes *C* and $C'$ when classifying *P*. For multi-class classification, SVMs can use the "one-against-one" classification system [4]. To decide whether or not *P* belongs to *C*, the system computes a score $S(P,C)$:

$$S(P,C) = \left|\{C' \neq C | f_{C,C'}(P) = C\}\right|$$

In other words, $S(P,C)$ is the number of classes $C'$ such that the SVM prefers *C* over $C'$ for classifying *P*. In the end, the element is classified to the class with the highest aggregate score. Therefore, *S* fits the definition of the matching function for SVMs: $S(P,C) = match(P,C)$.

The matching function of a classifier can be interpreted as its confidence. If $match(P,C)$ is low for all classes, the classifier is reluctant to classify *P* to any class. Normally, the classifier will nevertheless choose the highest-scoring class. This causes false positives despite the classifier's uncertainty. A confidence-based PO would instead classify the element as negative if its match is ambiguous ($match(P,C)$ is too low for all C).

Our confidence-based PO works as follows. Suppose that the classes are ordered from highest match to lowest, such that $C_1$ matches *P* the most (i.e. $C_1$ is the assumed class) , followed by $C_2$, and so on, until $C_{N+1}$. We first scale all *match* values linearly so that $match(P,C_1) = 1$ and $match(P,C_{N+1}) = 0$. For parameters *K* and $M_{match}$, we reject classification (classify the element as negative) if $\sum_{i=2}^{K+1} match(P,C_i) > K \cdot M_{match}$. In other words, we classify an element as negative if the top $K$ competing classes to the assumed class have a mean *match* score of $M_{match}$ or above. We vary $K$ and $M_{match}$ and test their effects on *r*-precision.

Table 5: Best 10-precision ($\pi_{10}$) and 1000-precision ($\pi_{1000}$) with confidence-based PO.

| Name | $\pi_{10}$ | $\pi_{1000}$ |
|---|---|---|
| Bi-XCor [1] | $.93 \pm .04$ | $.13 \pm .04$ |
| Pa-SVM [22] | $.306 \pm .009$ | $.0052 \pm .0002$ |
| Ca-OSAD [3] | $.77 \pm .08$ | $\geq .18$ |
| Wa-kNN [28] | $.91 \pm .03$ | $.11 \pm .02$ |
| Ha-kFP [9] | $\geq .97$ | $\geq .78$ |
| Pa-CUMUL [21] | $.43 \pm .02$ | $.0083 \pm .0005$ |

The output of *match* is also useful in cases where the attacker may want to rank his classifications, or explicitly output the confidence of classification. We expand on this use case in Section 4.1.

**Which attacks apply?**

All known attacks in the WF literature can be said to compute a *match*(*P,C*) function for all *C* and choosing the highest-scoring class to classify *P*. Therefore, the confidence-based PO applies to all of our classifiers.

**Results**

We present the results of confidence-based PO with regards to how it improves *r*-precision of our chosen WF attacks in Table 5.

With confidence-based PO, we see that Ha-kFP becomes highly precise even under the difficult 1000-precision scenario. In the optimal case ($K=2$, $M_{match} = 0.11$), we observed $R_{TP} = 0.318 \pm 0.004$, $R_{WP} = 0.0014 \pm 0.0003$, and $R_{FP} = 0$ (there were no false positives), giving $\pi_{1000} \geq 0.78$ as shown in the table. In fact, we found its mean precision to be much higher (0.996), but we do not use this unstable value as explained in Section 2.2.

The greatest improvement was observed in Bi-XCor, where 10-precision saw a 15-fold increase and 1000-precision saw a 200-fold increase. For Bi-XCor, we originally measured a surprising false positive rate of $R_{FP} = 0.89$, but this was reduced to $R_{FP} = 0.0014$ using our confidence-based PO. Bi-XCor without POs was unable to deal with the open-world scenario, though it was reasonably capable of handling the closed-world scenario; our methods bridged the gap between the two. On the other hand, the SVM-based classifiers Pa-SVM and Pa-CUMUL did not gain a significant improvement using confidence-based metrics. This may be because the *match* function we chose for SVMs was not sufficiently informative about classifier confidence.

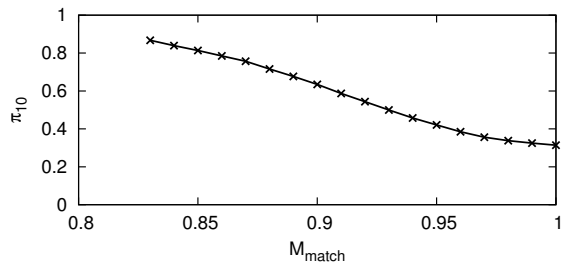

Figure 3: 10-precision with $K = 3$ and varying $M_{match}$ from 0.83 to 1.

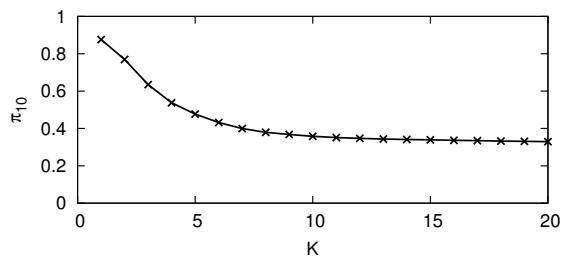

Figure 4: 10-precision with $M_{match} = 0.9$ and increasing $K$ from 1 to 20.

Optimal values for $K$ and $M_{match}$ differed for each algorithm. For the best-performing algorithm, Ha-kFP, the values were $K = 2$ and $M_{match} = 0.11$. It is interesting to note that all attacks were most precise when $K = 1$ or $K = 2$, i.e. it is optimal to compare the confidence of the assumed class only with the best competing class or the best two competing classes, and not more. On the other hand, the optimal value of $M_{match}$ is a consequence of the definition of *match* (detailed in the Abstract).

Ca-OSAD has a low lower bound in the $\pi_{1000}$ case because, like Ha-kFP, we were able to find a scenario in which there were no false positives; the TPR was about 20%. However, for Ca-OSAD specifically, due to computational restrictions, the number of non-monitored pages tested was 10 times smaller (5,000 instead of 50,000). This causes the precision to be unstable as shown by the Wilson method.

In Figure 3, we show the effects of decreasing $M_{match}$ on the confidence-based PO when applied to Wa-kNN. A lower $M_{match}$ represents a more cautious PO: the PO will reject the assumed class if the top $K$ classes (besides the assumed class) have a mean scaled *match* score above $M_{match}$. Consequently, $M_{match} = 1$ disables the PO. Here, we set $K = 3$, decrease $M_{match}$ from 1 to 0.83, and show the effects on 10-precision. (For $M_{match} < 0.83$, $R_{TP} < 0.2$ and therefore we rejected these results.) We see a significant improvement in 10-precision from $0.314 \pm 0.004$ to $0.867 \pm 0.003$ as we decrease $M_{match}$.

Similarly, in Figure 4, we increase $K$, the number of competing classes, from 1 to 20 and evaluate 10-precision. Here, we fix $M_{match} = 0.9$. An increased $K$ de-creases 10-precision; and optimal 10-precision is found at $K = 1$. This is expected as an increased $K$ leads to a lower average *match* and therefore the PO is less likely to reject the assumed class. Although this does not imply that optimal 10-precision will necessarily be found at $K = 1$ as we vary both $K$ and $M_{match}$, we did indeed find the optimal 10-precision and 1000-precision of Wa-kNN at $K = 1$ and $M_{match} = 0.88$.

Table 6: Best $\pi_{10}$ and $\pi_{1000}$ with the too-far PO and too-close PO on a 100x100+10000 data set.

| Name | Too-close PO | | | Too-far PO | | |
|---|---|---|---|---|---|---|
| | $\pi_{10}$ | $\pi_{1000}$ | Best distance | $\pi_{10}$ | $\pi_{1000}$ | Best Distance |
| Bi-XCor [1] | $.54 \pm .06$ | $.016 \pm .004$ | Bi-XCor | $.60 \pm .04$ | $.015 \pm .002$ | Pa-SVM |
| Pa-SVM [22] | $.96 \pm .08$ | $\geq .26$ | Bi-XCor | $.96 \pm .07$ | $\geq .22$ | Bi-XCor |
| Ca-OSAD [3] | $.93 \pm .09$ | $\geq .15$ | Bi-XCor | $.94 \pm .06$ | $.20 \pm .18$ | Pa-CUMUL |
| Wa-kNN [28] | $.95 \pm .07$ | $\geq .14$ | Bi-XCor | $.92 \pm .07$ | $.12 \pm .10$ | Bi-XCor |
| Ha-kFP [9] | $.96 \pm .08$ | $\geq .49$ | Bi-XCor | $.99 \pm .03$ | $\geq .61$ | Pa-SVM |
| Pa-CUMUL [21] | $.97 \pm .08$ | $\geq .18$ | Bi-XCor | $.96 \pm .07$ | $\geq .16$ | Bi-XCor |

### 3.3 Distance-based PO

Several WF attacks use or induce a notion of distance between packet sequences when performing classification.[6] We found that those distances, when used to augment the normal classification algorithm of WF attacks, could serve to remove questionable positive classifications and thus improve precision.

We derived distances between packet sequences based on known attacks. For example, we defined a distance based on Pa-SVM by executing its feature extraction algorithm, and then applying the radial basis function on the extracted features. From each distance between packet sequences, we derive a distance between packet sequences and classes. Note that while classifiers always chose the class with the highest *match* score, they did not choose the class with the shortest distance. For details on how we derived distances from WF attacks, we refer the reader to the Abstract. Then, we tested two different distance-based POs:

1. **Too-far PO**: We trained the PO by computing expected in-class distance (distance between packet sequences of the same class, for each class). If the distance of a testing packet sequence to the assumed class was more than $M_{distfar}$ times the expected in-class distance, we rejected classification.

2. **Too-close PO**: If there were at least $M_{distclose}$ classes that were closer to the packet sequence than its assumed class, we rejected classification.

---

[6]In our work, we do not use the strict mathematical definition of a "metric" when referring to distances. In particular, many of our distances do not satisfy the triangle inequality in edge cases. Rather, the distance quantifies the difference between packet sequences from the classifier's perspective. We avoid use of the word "metric" for this reason, opting to use the word "distance".

**Which attacks apply?**

We tested five distances, each one based on a different attack; Ha-kFP did not produce a distance. All classifiers can be optimized with both distance-based POs, even if it itself does not produce a distance. This means that we have a total of 60 optimized classifiers (two types of POs, five distances, six classifiers).

**Results**

We present the results for the two distance-based POs in Table 6. For both data sets, it was impossible in terms of both memory and computational time for us to compute the distance between all elements in the full 100x500+50000 data set. Instead, we reduced the data set to 100x100+10000 for both distance-based POs.

Interestingly, for the too-far PO, the best distance to use was always the Bi-XCor distance (the cross-correlation distance between packet timings and packet sizes). This distance contained significant information not sufficiently incorporated in future attacks, perhaps because it had achieved a comparatively low TPR. Appropriately, the distance for Bi-XCor could not save itself from relatively poor precision; no other distance could, either. This would suggest that Bi-XCor's distance is useful, but its classifier is weak (in fact, it only uses a naïve comparison between distances to classify). Poor results in the $\pi_{1000}$ scenario are due to the relatively small experiment size (100x100+10000).

The optimal value for $M$ was generally slightly less than 1 for each of the above classifiers. A larger $M$ weakened the precision optimizer, but a smaller $M$ may cause TPR to drop too significantly. We show the effect of $M$ on the lower bound of $\pi_{1000}$ for Pa-CUMUL and the Bi-XCor distance in Figure 5. Peak precision occurred at $M = 0.925$. It makes sense that $M > 1$ would be imprecise: this was the point after which the PO would not reject classification even if the distance to its assumed class was greater than expected. There was almost no change in precision beyond $M = 2$ (when the PO almost never rejected a classification). The figure also shows the somewhat undesirable result that $\pi_{1000}$ is predicate on choosing a correct $M$.

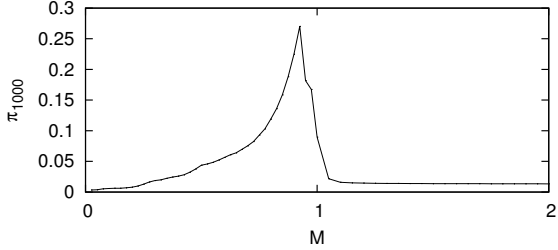

Figure 5: Lower bound of $\pi_{1000}$ for `Pa-CUMUL` with too-far PO, using the `Bi-XCor` distance on 100x100+10000 elements while varying $M$. The PO rejected any assumed class for which the testing element had a distance that was at least $M$ times the expected distance to that class.

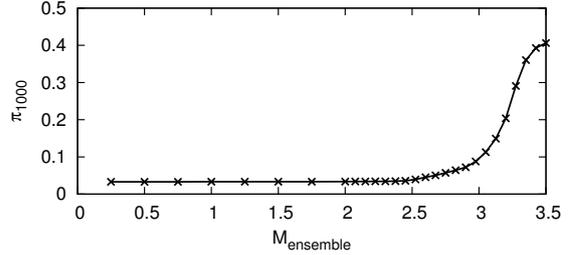

Figure 6: Lower bound of $\pi_{1000}$ for the match-based ensemble PO when varying $M_{ensemble}$, the minimum sum match score, below which we would reject classification. With $M_{ensemble} > 3.5$, $R_{TP} < 0.2$ and thus we discarded these results.

For the too-close PO, unlike the too-far PO, it was not always the case that the `Bi-XCor` distance served best to optimize precision. We show the best distance to use in each column as well. The performance of most attacks was similarly improved under the too-close PO compared to the too-far PO. `Ha-kFP` was able to achieve a large increase in precision in the $\pi_{1000}$ case.

For `Ca-OSAD`, the optimal $M$ value to use was 2; for all others, it was 1. $M = 1$ means that the PO rejected the classification of any element that had a smaller distance to a class that was not the classifier's assumed class. This is similar (though not identical) to an ensemble scheme where we ask two classifiers to classify each testing element, and rejected any classification where the outputs were different. The good performance of the above scheme suggests that ensemble learning may potentially lead to better-optimized classifiers.

### 3.4 Ensemble PO

In ensemble learning, multiple classifiers simultaneously classify the same testing element, and we decide the final class based on an aggregate of each classifier's individual classification. In adapting ensemble learning techniques for OWF, we hypothesized that disagreements between different classifiers could show a lack of confidence. Therefore, we should reject classification when different classifiers output different classifications.

We first evaluate a simple bagging scheme for OWF. We trained all classifiers except `Ca-OSAD` (as it could not be trained on the full data set) separately on the same training set using 10-fold classification. Then, we take a subset of the classifiers and, for each testing element, we ask each of them to determine the assumed class. We rejected classification whenever there was no unanimous decision among all classifiers in the chosen subset of classifiers. Therefore, the more classifiers there are, the more conservative our classifications become.

We show the results for all 31 possible subsets of 5 classifiers in Table 7. We only show the minimum $\pi_{1000}$ (Wilson method) due to space. The black marker indicates which algorithms are used. Filled black rows represent that the algorithm forms part of the subset for that result. From top to bottom, they represent `Bi-XCor`, `Pa-SVM`, `Wa-kNN`, `Ha-kFP`, and `Pa-CUMUL`. Row $i$ contains results for the use of $i$ classifiers in ensemble. For example, the second result in the third row represents the use of `Bi-XCor`, `Pa-SVM` and `Ha-kFP`. All results here exceed a recall of 0.2.

From Table 7, we see that the optimal precision is achieved when all five classifiers are used for $\pi_{1000} \geq 0.718$. We observed no false positives in this case, i.e. no element tricked all five classifiers into thinking it belonged to the same sensitive page. The best single classifier was `Ha-kFP`; the best two-classifier ensemble was to add `Wa-kNN`; the best three-classifier ensemble was to add `Pa-SVM`; and the best four-classifier ensemble was to add `Pa-CUMUL`. Any ensemble that included `Ha-kFP` was especially precise; for example, the worst four-classifier ensemble was the one that did not include `Ha-kFP`. Using more classifiers made classification more conservative and gave better precision.

We also tested a scheme where all classifiers are used: we first determine a combined assumed class by asking each classifier individually what their assumed class is, and choosing the most popular class. (Ties are broken by more precise classifiers based on Table 4.) Then, if the sum *match* score of all classifiers for that class is less than some parameter $M_{ensemble}$, we reject classification. We show the minimum $\pi_{1000}$ (Wilson method) results of this scheme in Figure 6. Unfortunately, though this scheme seems more sophisticated than the above, its $\pi_{1000}$ did not reach the high levels as the trivial all-classifiers ensemble PO above; the peak was $\pi_{1000} \geq 0.406$. Results are somewhat sensitive to a correct choice of $M_{ensemble}$, as precision quadruples from $M_{ensemble} = 3$ and $M_{ensemble} = 3.5$. Beyond $M_{ensemble} = 3.5$, recall dropped below 0.2, so we only accepted results for $M_{ensemble} \leq 3.5$.

Table 7: Lower bound for $\pi_{1000}$ when a subset of the five WF attacks are used in ensemble PO. The marker besides each result indicates which of the five WF attacks are used (black bar = used, white bar = not used). From top to bottom, the bars represent Bi-XCor, Pa-SVM, Wa-kNN, Ha-kFP, and Pa-CUMUL in order.

| | | | | | | | | | |
|---|---|---|---|---|---|---|---|---|---|
| .007 | .004 | .005 | .013 | .005 | | | | | |
| .039 | .032 | .059 | .038 | .065 | .173 | .056 | .225 | .049 | .131 |
| .228 | .447 | .205 | .373 | .158 | .376 | .543 | .200 | .387 | .438 |
| .656 | .362 | .590 | .590 | .705 | | | | | |
| .718 | | | | | | | | | |

The above scheme implicitly assigns equal weights to each classifier's vote (except for the tie-breaking routine). This seems to defy our previous intuition that, for example, Ha-kFP should be weighed more heavily. Therefore, we attempted to multiply weights between to each classifier's *match* score before adding them together to determine if they exceed some $M_{ensemble}$. As there are now six parameters, the search space is relatively large; before developing a procedure to learn parameters, we decided to exhaustively search the parameter space to see if any parameters gave good results. (defying the ML rule that training should not be used for testing).

We searched randomly for weights between 0 and 1 so that the sum weight is equal to 1, and we searched for 24 hours to find 40,000 possible combinations of weights. The best result we found assigned weights $\langle 0.028, 0.342, 0.005, 0.249, 0.376 \rangle$ to the five algorithms with $M = 0.9$ to achieve $\pi_{1000} \geq 0.663$. As this is less promising than the simpler ensemble scheme shown in Table 7, where we had achieved $\pi_{1000} \geq .718$, we did not test a parameter learning algorithm, and we do not recommend using this parameter-sensitive algorithm.

## 4 Uses for a precise WF attack

### 4.1 Selection

We consider a scenario where the attacker knows that a single sensitive web page was accessed at some point, and wants to know who did so. With auxiliary knowledge (for example, when the sensitive access took place), the attacker has narrowed it down to $S$ particular candidate accesses: one is sensitive and all the others are non-sensitive. For example, such an attacker could be someone who wants to track who leaked a particular sensitive document on Tor, in which case he would look for an access to the leaking website, and he would be able to narrow down to a number of candidates based on when the leak took place.

To test if a precise WF attack would be useful in this scenario, we set up an experiment to compare the same classifier with PO and without PO. Randomly, we choose one out of the sensitive web pages and access it once; then, we make $S-1$ accesses to non-sensitive web pages.

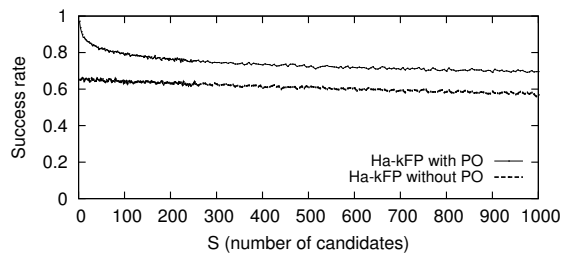

Figure 7: Success rate of identifying a sensitive access out of $S$ possible candidates for Ha-kFP with and without the confidence-based PO.

These accesses were taken randomly from the same data set for previous experiments. The classifier with PO selects the access with the highest *match* score as the sensitive web page access. The classifier without PO tries to classify all elements; if multiple accesses are classified to the sensitive web page, the classifier chooses one randomly. We repeat the above for 10,000 trials for each $S$ and take the mean of the success rate.

We compare Ha-kFP with and without confidence-based PO in Figure 7. When there are 100 candidates, Ha-kFP with PO has a 80% chance of identifying the correct page access, while Ha-kFP without PO has a 63% chance. The success rate of Ha-kFP without PO comes mostly from its TPR; while Ha-kFP with PO has a lower TPR, its greater success in distinguishing the sensitive site comes from its use of the *match* score. For both algorithms, the decay in success rate is slow when $S$ increases — the random guess success rate would be $1/S$.

### 4.2 Identifying a sensitive client

We want to know if the attacker could determine, after some period of observing the client, whether or not the client has a habit of visiting a particular sensitive page. This scenario simulates an attacker who wants to identify the client's online behavior. For example, the attacker may want to figure out the client's political affiliation, romantic status, or other demographics by deciding if the client visits certain pages frequently.

Let us define a sensitive client as one who visits a sensitive page at a rate of $b$, and a non-sensitive client as one who does not visit that sensitive page. The attacker's goal is to distinguish sensitive clients and non-sensitive clients for some $b$. To do so, the attacker observes each client for $N$ page accesses, performs some WF classification on their accesses, and gets some number $x$ of sensitive page accesses. The attacker decides that the client is a sensitive client if $x > M_{identify}$. $M_{identify}$ is a parameter that controls the trade-off between identifying sensitive clients correctly and mistaking non-sensitive clients as sensitive clients.

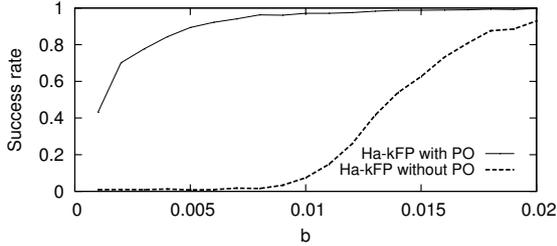

Figure 8: Success rate of identifying a sensitive client for `Ha-kFP` with and without confidence-based PO, while varying $b$, the rate of sensitive page visits.

We compare `Ha-kFP` with and without confidence-based PO in Figure 8. We vary the base rate between 0.001 and 0.02 and model an attacker who has observed 1000 page accesses. Therefore, in the toughest $b = 0.001$ case, the client has only visited the sensitive page once. We set $M_{identify}$ so that the rate of mis-identifying non-sensitive clients as sensitive clients was about 1% for both algorithms. (This was $M_{identify} = 0$ for `Ha-kFP` with PO and $M_{identify} = 8$ for `Ha-kFP` without PO). We can see that the confidence-based PO is much more successful at identifying sensitive clients when $b$ is low, with a 70% chance of correct identification at $b = 0.002$, up to 99% at $b = 0.014$.

It is easier to identify sensitive clients if they are monitored for a longer period of time. If the attacker observed 2000 page accesses, he would have a 86% chance of identifying a client that visits a sensitive page at a base rate of 0.002. This is a disadvantage of Tor's long-term guard policy; currently, Tor clients access a pool of three guards, which last between 30 to 60 days [8], and it has been proposed that they should last for 9 months [6]. Tor guards are potential WF attackers. For this specific scenario, it would perhaps be better for privacy if Tor guards had a small portion of every client's traffic, rather than the full picture of a small number of clients' traffic.

### 4.3 Attacking Defenses

A number of defenses against WF have been proposed for anonymity technologies like Tor. Much like for WF attacks, these defenses are almost always evaluated with recall: a good defense would be judged by its ability to decrease the recall of all classifiers. We wanted to know the precision of our optimized attacks on those defenses and whether or not their precision could be significantly deterred by defenses.

We evaluated three WF defenses: Random Padding set to produce 50% bandwidth and time overhead by adding random packets, Tamaraw by Cai et al. [2], and Adaptive Padding by Juarez et al. [13]. Bandwidth overhead is equal to the percentage of extra packets added by the defense, and time overhead is equal to the percentage of extra time required to load each web page.

We present the results in Table 8. For these results only, we lower the minimum recall rate requirement from 0.2 to 0.02, because we chose settings for which no attack can achieve a higher recall than 0.06 for Tamaraw. For Adaptive Padding, we could only test a smaller 100x50+5000 data set because it is very slow in computation; the full experiment would have taken a month. We can see that Adaptive Padding is better in every metric compared to Random Padding, though $\pi_{10}$ can be as high as 0.25 for an attacker. Even without any PO, attacking Adaptive Padding was relatively easy compared to Tamaraw. However, Tamaraw suffers a much higher time overhead; while it is possible to perform a trade-off between bandwidth and time overhead [2], Tamaraw cannot achieve overhead values close to those of Adaptive Padding. In addition, a large jump in precision is possible in Tamaraw with confidence-based PO. Unlike Adaptive Padding, Tamaraw has a guaranteed maximum TPR for any WF attack (depending on the parameters), but such a guaranteed maximum TPR does not lead to a guaranteed minimum precision.

We did not evaluate any "targeted defenses": a targeted defense allows the client to choose which page to mimic while accessing specific pages. While targeted defenses give clients the advantage of creating a specific cover story, rather than a series of randomly perturbed packet sequences, currently there is no known mechanism to automatically choose correct targets to mimic. Wrongly chosen targets could significantly impede the ability of the defense to lower precision, so we cannot evaluate them fairly. These defenses include Glove [20], Walkie-Talkie [31], and Decoy pages [22].

## 5 Discussion

### 5.1 Does Website Fingerprinting scale?

The size of our open world (50,000 pages) is unprecedented in WF literature, and we have shown that optimized classifiers can be highly precise in our large open world. Yet, the reader may argue that our experimental open world is tiny compared to the actual open world (with over a billion pages). Can these results be extended to the actual open world?

Table 8: Best 10-precision ($\pi_{10}$) with confidence-based PO on Wa-kNN against three defenses: Random Padding, Tamaraw and Adaptive Padding.

| Name | Original | | With PO | | Overhead | |
|---|---|---|---|---|---|---|
| | TPR | $\pi_{10}$ | TPR | $\pi_{10}$ | Bandwidth | Time |
| Random Padding | .413 ± .004 | .093 ± .002 | .021 ± .001 | .446 ± .076 | 50% | 50% |
| Tamaraw [2] | .053 ± .002 | .0065 ± .0002 | .022 ± .001 | .129 ± .001 | 58% | 162% |
| Adaptive Padding [13] | .27 ± .01 | .076 ± .006 | .029 ± .001 | .25 ± .11 | 32% | 0% |

So long as the experimental procedure is correct, the effectiveness of a system does not have to be evaluated on all potential inputs. The essential procedure in a correct open-world experiment is that **the classifier should never be tested on the same non-sensitive pages it has trained on**. As long as we follow this rule, the real open world size does not matter, as the attacker is (pessimistically) assumed to have no knowledge about any non-sensitive pages in the open world testing set.

To achieve this experimentally, we visited each non-sensitive page only once, and split the training and testing set (generally using 10-fold cross validation, though the training set was smaller for some attacks). We also filtered out pages with similar domain names so that they would not both appear in our non-monitored set. Since the classifier has no knowledge whatsoever of the non-sensitive pages in the testing set, its decision that a testing element should be classified as non-sensitive is not based on any specific knowledge about the page it came from. In other words, the actual open world — and its larger size — has no impact on the classifier's success.

The above rule leads to a conservative measure of a WF attack's success (recall and precision). A client can visit one of three pages: a sensitive page, a non-sensitive page that the attacker did not train on, and a non-sensitive page that he did train on. The third case is easier than the second case, but we do not allow it to happen in our experiments; it may happen in a realistic scenario when the attacker trains on a large open world.

As a comparison, we observe that while there may be billions of humans, a DNA recognition or fingerprint recognition classifier can still be considered successful without evaluating the classifier on the majority of humans. For the same reason, it is important to evaluate the classifier based on precision for low base-rate scenarios.

### 5.2 Is Website Fingerprinting realistic?

Generally, WF works make several assumptions to present results. These assumptions include the use of a cold cache, freshness of the training set and the lack of noise packets unrelated to the page that could impede classification. Several authors [12, 30] have noted that violating these assumptions would cause recall rate to drop. We note that assuming the use of a cold cache is reasonable for Tor (as it does not keep cache in disk), and Wang et al. [30] have shown that the training set can be kept sufficiently fresh for classification. Wang et al. also showed that classification is only impeded when the bandwidth rate of noise is very high, possibly from a video in a separate tab.

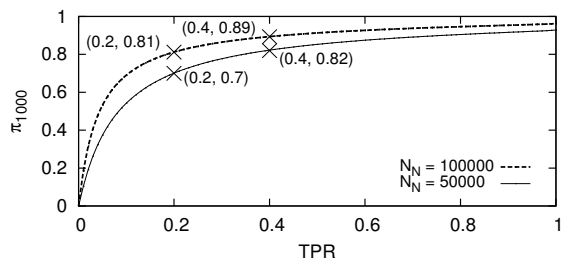

Figure 9: Lower bound for $\pi_{1000}$ by the Wilson method when both WPR and FPR are set to 0, and TPR is varied. $N_N = 50000$ indicates the use of 50,000 non-sensitive pages, as is the case in our experiments; $N_N = 100000$ represents what would happen if we had used 100,000 non-sensitive pages instead.

Whether or not the above problems are solved, WF attacks must certainly have high precision to succeed in the realistic open world, which this work focuses on.

### 5.3 The way forward for $r$-precision

The reader may have noticed that some of our POs can give classifiers a very low false positive rate (even zero), but $\pi_{1000}$ does not exceed 0.8 using the Wilson method in our experiments. Even when there are no false positives (in which case $\pi_{1000}$ should be 1), the Wilson method correctly acknowledges that it is not likely that a separate data set would also have no false positives, and does not output $\pi_{1000} = 1$.

To determine how $\pi_{1000}$ can be improved further, in Figure 9, we show the lower bound for $\pi_{1000}$ by varying TPR and setting both WPR and FPR to 0. We do this for two cases, one matching our experimental scenario with 50,000 non-sensitive pages, and one for a different scenario with 100,000 non-sensitive pages.

We see that it is possible to improve $\pi_{1000}$ substantially with higher TPR. If we had achieved a TPR of 0.4, WPR of 0 and FPR of 0 in our experiments, then we would have $\pi_{1000} \geq 0.82$. A "perfect" result where TPR is 1 would have $\pi_{1000} \geq 0.93$, showing the conservative nature of the Wilson method. Alternatively, it is possible to improve $\pi_{1000}$ by performing an experiment with more non-sensitive pages. If we had achieved the same TPR of 0.2 on a larger data set with 100,000 non-sensitive pages, we would have $\pi_{1000} \geq 0.81$. This shows the intuitively correct notion that increasing the experiment size should improve the results of a conservative estimate.

# 6 Related Work

Traffic analysis attacks are a threat to web browsing privacy. Cheng et al. [5], Sun et al. [25], Hintz et al. [11], and Bissias et al. [1] were some of the first to show successful classifiers to determine which page someone is visiting based on traffic patterns. Later works referred to this traffic analysis problem as website fingerprinting.

The original paper on Tor considered traffic analysis to be a serious threat [7], though no attack had been successful on Tor at that time as it equalized cell sizes. In particular, Herrmann et al. [10] showed that their attack, as well as Liberatore and Levine's attack [15], did not succeed against Tor; the authors were able to beat SSH and some VPNs. Lu et al. [16], Panchenko et al. [22] and Cai et al. [3] were some of the first to show success against Tor. Attack accuracy was improved and computational time was decreased by Wang et al. [28], Hayes et al. [9] and another work by Panchenko et al. [21]. These works also began to note the significance of the open world and conduct experiments on them.

Some previous WF works have discussed the base rate fallacy, though no explicit attempt has been made to resolve it. Panchenko et al. [21] discussed issues with precision, though their attack was not precise (as seen in our results as well), and they found that WF attacks generally would fail in the large open world if they are not precise. This important point underlines our paper's motivation. Wang et al. [28] noted that an increase in $k$ for their $k$-NN classifier decreased both TPR and FPR; though they did not evaluate precision, an increase in $k$ in their algorithm can be considered an implicit attempt to optimize precision for a WF attack.

# 7 Conclusion and Future Work

This work solves the open problem of open-world website fingerprinting (OWF). We argue that precision is more important than recall for OWF, which suggests that previous classifiers were not correctly designed for OWF. To resolve this issue, we formulate $r$-precision ($\pi_r$), which is defined as the percentage of sensitive classifications that are correct if the client visits $r$ times as many non-sensitive pages as sensitive pages.

We present three classes of POs to improve the precision of WF classifiers. Our confidence-based POs (based on a *match* score) ask classifiers to output the degree of confidence they have in their classifications, and we reject classifications that are not confident enough. When applied to `Ha-kFP`, this produced $\pi_{1000} \geq 0.78$. Our distance-based POs reject classifications of testing elements that are too far from the assumed class. This produced $\pi_{1000} \geq 0.61$ with a distance based on `Pa-SVM` when applied to `Ha-kFP`. Our ensemble-based POs reject elements for which a chosen set of classifiers did not unanimously agree on the assigned class. When all five classifiers are used in ensemble, this produced $\pi_{1000} \geq 0.72$. Without POs, the best classifier only had a precision of $\pi_{1000} = 0.014$.

Our approach has a major advantage of being adaptable to any classifier designed for the closed world, allowing us to evaluate their effects on the breadth of previously proposed classifiers for the WF problem. Our high precision is the result of leveraging classifiers that are already well-built for the closed-world WF problem. However, this also poses a limitation: the classifier itself is not designed with precision in mind. A classifier designed from the start to optimize precision could perform better than our POs.

Previous authors have shown that a number of problems in realistic scenarios would lower recall, such as noisy packet sequences, poor training sets and multi-tab browsing; solutions have been proposed in previous works. We do not know how these realistic problems and solutions would affect precision. We believe that a final way to demonstrate the practicality of OWF would be to create a private Tor guard that performed website fingerprinting on consenting Tor clients, telling them which sensitive web pages they visited and asking them if it is correct. Our optimized classifiers are up to the task.

On the flip side, defenses should also be designed to minimize the attacker's precision. One approach to designing defenses is to create "anonymity sets" of web pages that look the same to the attacker after padding. This approach gives guarantees on the maximum recall of any attacker; they may be extended to guarantee maximum precision when base rates are considered during design.

Our data and code can be found at:

https://github.com/OpenWF/openwf.git

**Acknowledgments**

Acknowledgments are scrubbed for the blind review submission.

# Appendix

Here we describe each of the six previously published attacks we tested with our POs. The attacks are Bi-XCor [1], Pa-SVM [22], Ca-OSAD [3], Wa-kNN [28], Ha-kFP [9] and Pa-CUMUL [21]. We describe how each classifier represents packet sequences $P$ as $R(P)$, the distance $d(P, P')$ between two packet sequences $P$ and $P'$, the training and the testing procedures. We describe the testing procedure by specifying *match* (as explained in Section 3.2); each classifier assigns the element to the class that scores the highest with *match*.

We denote packet sequences as $P = \langle p_1, p_2, \ldots, p_n \rangle$, where $p_i = (t_i, \ell_i)$, $t_i$ is the interpacket time between $p_{i-1}$ and $p_i$, and $\ell_i$ is the byte length of packet $p_i$, with positive packet lengths representing outgoing packets from the client and negative packet lengths representing incoming packets to the client. With Tor cells, $\ell_i \in \{-1, 1\}$ as all cells have the same size. $P$ represents the WF attacker's information, and he attempts to deduce which web page it came from.

## Bi-XCor

*Representation.* We split $R(P) = (R_t(P), R_\ell(P))$, where:

$$R_t(P) = \langle t_1, t_2, \ldots, t_n \rangle$$
$$R_\ell(P) = \langle \ell_1, \ell_2, \ldots, \ell_n \rangle$$

*Distance.* Consider two lists $a$ and $b$ with mean $\bar{a}$, $\bar{b}$ and standard deviation $\sigma_a$, $\sigma_b$ respectively. We define the cross-correlation function $X(a,b)$ between them:

$$X(a,b) = \frac{\sum_{i=1}^{\min(|a|,|b|)}(a_i - \bar{a})(b_i - \bar{b})}{\min(|a|,|b|) \cdot \sigma_a \cdot \sigma_b}$$

We have:

$$d(P, P') = 2 - X(R_t(P), R_t(P')) - X(R_\ell(P), R_\ell(P'))$$

*Training.* We represent each class $C$ as $R(C) = (R_t(C), R_\ell(C))$, where the $i$-th element of $R_t(C)$ is the mean of all $t_i$ for training packet sequences from class $C$, and similarly for $R_\ell(C)$.

*Testing.*

$$match(P, C) = d(R(P), R(C))$$

## Pa-SVM

*Representation.* We extract a number of features from each packet sequence related to packet ordering, directions, and sizes: $R(P) = <f_1, f_2, \ldots, f_{|F|}>$. To see the list of features, refer to the original work [22] or our code.

*Distance.* We use the radial basis function with $\gamma = 2^{-25}$ to compute distances between the feature representations of packet sequences. The distance is:

$$d(P, P') = 1 - e^{-\gamma ||R(P) - R(P')||^2}$$

*Training.* We train an SVM on the above pairwise distances by finding support vectors which separate classes.

*Testing.* The matching function uses one-against-one SVM classification as described in Section 3.2.

## Ca-OSAD

*Representation.*

$$R(P) = \{\ell_1, \ell_2, \ldots\}$$

*Distance.* We compute the pairwise distance between packet sequences $P$ and $P'$ as:

$$d(P, P') = 1 - e^{-2 \cdot OSAD(P,P')^2 / \min(|P|, |P'|)}$$

In the above, $OSAD(P, P')$ is the Optimal String Alignment Distance between $R(P)$ and $R(P')$.

*Training.* We train an SVM using the custom kernel calculated from the above pairwise distances.

*Testing.* The matching function uses one-against-one SVM classification as described in Section 3.2.

## Wa-kNN

*Representation.* We extract a number of features from each packet sequence related to packet ordering, directions, and sizes: $R(P) = <f_1, f_2, \ldots, f_{|F|}>$. To see the list of features, refer to the original work [28] or our code.

*Distance.* We use a weighted $L_1$-distance between $P$ and $P'$:

$$d(P, P') = \sum_{i=1}^{|F|} w_i |f_i - f'_i|$$

*Training.* We learn weights $w_i$ that optimize the accuracy of the weighted distance.

*Testing.*

$$match(P, C) = \min_{P' \in C} d(P, P')$$

## Ha-kFP

*Representation.* We extract features from each packet sequence, similar to Wa-kNN. To see the list of features, refer to the original work [9] or our code.

*Distance.* Ha-kFP does not produce a distance.

*Training.* We train a Random Forest classifier with 1000 decision trees, where each tree draws a random sample of the input elements with replacement, resulting in a sample of equal size to the input. Each leaf $L$ of a decision tree records $L(x)$, the number of training samples of each class that fell in that leaf, for class $x$.

*Testing.*

If $P$ falls in leaf $L$ for decision tree $i$, we calculate $match_i(P,C) = L(C)/\sum_x L(x)$. Then

$$match(P,C) = \sum_{i=1}^{1000} match_i(P,C)$$

**Pa-CUMUL**

*Representation.* We extract features from each packet sequence, based on total size, time, and 100 linear interpolations of aggregated packet sizes. To see the list of features, refer to the original work [21] or our code.

*Distance.* We use the radial basis function with $\gamma = 2^{-28}$ to compute distances between the feature representations of packet sequences. The distance is:

$$d(P,P') = 1 - e^{-\gamma||R(P)-R(P')||^2}$$

*Training.* We train an SVM on the above pairwise distances by finding support vectors which separate classes.

*Testing.* The matching function uses one-against-one SVM classification as described in Section 3.2.

**Distances**

For our distance-based POs, we derived a distance between packet sequences $P, P'$ based on five previous WF attacks: `Bi-XCor`, `Pa-SVM`, `Ca-OSAD`, `Wa-kNN` and `Pa-CUMUL`. The distance is equivalent to $d(P,P')$ as written above for each WF attack. Then, we derived a distance between packet sequences $P$ and classes $C$ based on the distance between packet sequences as follows. We denote $C[:N]$ to mean the $N$ closest elements to $P$ in $C$, and $C[N]$ to mean the $N$-th closest element to $P$ in $C$.

1. $d(P,C) = \sum_{P' \in C} d(P,P')/|C|$.
2. $d(P,C) = \sum_{P' \in C[:5]} d(P,P')/|C|$.
3. $d(P,C) = \sum_{P' \in C[:25]} d(P,P')/|C|$.
4. $d(P,C) = d(P,C[1])$.
5. $d(P,C) = d(P,C[5])$.
6. $d(P,C) = d(P,C[25])$.